\documentclass[twocolumn,letterpaper,aps,prl,superscriptaddress,showpacs,floatfix]{revtex4-1}
\usepackage{graphicx}	% Include figure filed

\newcommand{\lt}{<}

\begin{document}

%Title of paper

\title{Comparison of the space-time extent of the emission source in 
$d$$+$Au and Au$+$Au collisions at $\sqrt{s_{{NN}}}$ = 200 GeV}

\newcommand{\abilene}{Abilene Christian University, Abilene, Texas 79699, USA}
\newcommand{\augie}{Department of Physics, Augustana College, Sioux Falls, South Dakota 57197, USA}
\newcommand{\banaras}{Department of Physics, Banaras Hindu University, Varanasi 221005, India}
\newcommand{\barc}{Bhabha Atomic Research Centre, Bombay 400 085, India}
\newcommand{\baruch}{Baruch College, City University of New York, New York, New York, 10010 USA}
\newcommand{\bnlcoll}{Collider-Accelerator Department, Brookhaven National Laboratory, Upton, New York 11973-5000, USA}
\newcommand{\bnlphys}{Physics Department, Brookhaven National Laboratory, Upton, New York 11973-5000, USA}
\newcommand{\caucr}{University of California - Riverside, Riverside, California 92521, USA}
\newcommand{\charlesczech}{Charles University, Ovocn\'{y} trh 5, Praha 1, 116 36, Prague, Czech Republic}
\newcommand{\chonbuk}{Chonbuk National University, Jeonju, 561-756, Korea}
\newcommand{\ciae}{Science and Technology on Nuclear Data Laboratory, China Institute of Atomic Energy, Beijing 102413, P.~R.~China}
\newcommand{\cns}{Center for Nuclear Study, Graduate School of Science, University of Tokyo, 7-3-1 Hongo, Bunkyo, Tokyo 113-0033, Japan}
\newcommand{\colorado}{University of Colorado, Boulder, Colorado 80309, USA}
\newcommand{\columbia}{Columbia University, New York, New York 10027 and Nevis Laboratories, Irvington, New York 10533, USA}
\newcommand{\czechtech}{Czech Technical University, Zikova 4, 166 36 Prague 6, Czech Republic}
\newcommand{\dapnia}{Dapnia, CEA Saclay, F-91191, Gif-sur-Yvette, France}
\newcommand{\debrecen}{Debrecen University, H-4010 Debrecen, Egyetem t{\'e}r 1, Hungary}
\newcommand{\elte}{ELTE, E{\"o}tv{\"o}s Lor{\'a}nd University, H - 1117 Budapest, P{\'a}zm{\'a}ny P. s. 1/A, Hungary}
\newcommand{\ewha}{Ewha Womans University, Seoul 120-750, Korea}
\newcommand{\fit}{Florida Institute of Technology, Melbourne, Florida 32901, USA}
\newcommand{\fsu}{Florida State University, Tallahassee, Florida 32306, USA}
\newcommand{\gsu}{Georgia State University, Atlanta, Georgia 30303, USA}
\newcommand{\hanyang}{Hanyang University, Seoul 133-792, Korea}
\newcommand{\hiroshima}{Hiroshima University, Kagamiyama, Higashi-Hiroshima 739-8526, Japan}
\newcommand{\howard}{Department of Physics and Astronomy, Howard University, Washington, DC 20059, USA}
\newcommand{\ihepprot}{IHEP Protvino, State Research Center of Russian Federation, Institute for High Energy Physics, Protvino, 142281, Russia}
\newcommand{\illuiuc}{University of Illinois at Urbana-Champaign, Urbana, Illinois 61801, USA}
\newcommand{\inrras}{Institute for Nuclear Research of the Russian Academy of Sciences, prospekt 60-letiya Oktyabrya 7a, Moscow 117312, Russia}
\newcommand{\instpasczech}{Institute of Physics, Academy of Sciences of the Czech Republic, Na Slovance 2, 182 21 Prague 8, Czech Republic}
\newcommand{\isu}{Iowa State University, Ames, Iowa 50011, USA}
\newcommand{\jaea}{Advanced Science Research Center, Japan Atomic Energy Agency, 2-4 Shirakata Shirane, Tokai-mura, Naka-gun, Ibaraki-ken 319-1195, Japan}
\newcommand{\jinrdubna}{Joint Institute for Nuclear Research, 141980 Dubna, Moscow Region, Russia}
\newcommand{\jyvaskyla}{Helsinki Institute of Physics and University of Jyv{\"a}skyl{\"a}, P.O.Box 35, FI-40014 Jyv{\"a}skyl{\"a}, Finland}
\newcommand{\kek}{KEK, High Energy Accelerator Research Organization, Tsukuba, Ibaraki 305-0801, Japan}
\newcommand{\korea}{Korea University, Seoul, 136-701, Korea}
\newcommand{\kurchatov}{Russian Research Center ``Kurchatov Institute", Moscow, 123098 Russia}
\newcommand{\kyoto}{Kyoto University, Kyoto 606-8502, Japan}
\newcommand{\labllr}{Laboratoire Leprince-Ringuet, Ecole Polytechnique, CNRS-IN2P3, Route de Saclay, F-91128, Palaiseau, France}
\newcommand{\lahorelums}{Physics Department, Lahore University of Management Sciences, Lahore, Pakistan}
\newcommand{\lawllnl}{Lawrence Livermore National Laboratory, Livermore, California 94550, USA}
\newcommand{\losalamos}{Los Alamos National Laboratory, Los Alamos, New Mexico 87545, USA}
\newcommand{\lpc}{LPC, Universit{\'e} Blaise Pascal, CNRS-IN2P3, Clermont-Fd, 63177 Aubiere Cedex, France}
\newcommand{\lund}{Department of Physics, Lund University, Box 118, SE-221 00 Lund, Sweden}
\newcommand{\maryland}{University of Maryland, College Park, Maryland 20742, USA}
\newcommand{\mass}{Department of Physics, University of Massachusetts, Amherst, Massachusetts 01003-9337, USA }
\newcommand{\michigan}{Department of Physics, University of Michigan, Ann Arbor, Michigan 48109-1040, USA}
\newcommand{\muenster}{Institut fur Kernphysik, University of Muenster, D-48149 Muenster, Germany}
\newcommand{\muhlenberg}{Muhlenberg College, Allentown, Pennsylvania 18104-5586, USA}
\newcommand{\myongji}{Myongji University, Yongin, Kyonggido 449-728, Korea}
\newcommand{\nagasaki}{Nagasaki Institute of Applied Science, Nagasaki-shi, Nagasaki 851-0193, Japan}
\newcommand{\newmex}{University of New Mexico, Albuquerque, New Mexico 87131, USA }
\newcommand{\nmsu}{New Mexico State University, Las Cruces, New Mexico 88003, USA}
\newcommand{\ohio}{Department of Physics and Astronomy, Ohio University, Athens, Ohio 45701, USA}
\newcommand{\ornl}{Oak Ridge National Laboratory, Oak Ridge, Tennessee 37831, USA}
\newcommand{\orsay}{IPN-Orsay, Universite Paris Sud, CNRS-IN2P3, BP1, F-91406, Orsay, France}
\newcommand{\peking}{Peking University, Beijing 100871, P.~R.~China}
\newcommand{\pnpi}{PNPI, Petersburg Nuclear Physics Institute, Gatchina, Leningrad region, 188300, Russia}
\newcommand{\riken}{RIKEN Nishina Center for Accelerator-Based Science, Wako, Saitama 351-0198, Japan}
\newcommand{\rikjrbrc}{RIKEN BNL Research Center, Brookhaven National Laboratory, Upton, New York 11973-5000, USA}
\newcommand{\rikkyo}{Physics Department, Rikkyo University, 3-34-1 Nishi-Ikebukuro, Toshima, Tokyo 171-8501, Japan}
\newcommand{\saispbstu}{Saint Petersburg State Polytechnic University, St. Petersburg, 195251 Russia}
\newcommand{\saopaulo}{Universidade de S{\~a}o Paulo, Instituto de F\'{\i}sica, Caixa Postal 66318, S{\~a}o Paulo CEP05315-970, Brazil}
\newcommand{\seoulnat}{Department of Physics and Astronomy, Seoul National University, Seoul, Korea}
\newcommand{\stonybrkc}{Chemistry Department, Stony Brook University, SUNY, Stony Brook, New York 11794-3400, USA}
\newcommand{\stonycrkp}{Department of Physics and Astronomy, Stony Brook University, SUNY, Stony Brook, New York 11794-3800, USA}
\newcommand{\tenn}{University of Tennessee, Knoxville, Tennessee 37996, USA}
\newcommand{\titech}{Department of Physics, Tokyo Institute of Technology, Oh-okayama, Meguro, Tokyo 152-8551, Japan}
\newcommand{\tsukuba}{Institute of Physics, University of Tsukuba, Tsukuba, Ibaraki 305, Japan}
\newcommand{\vandy}{Vanderbilt University, Nashville, Tennessee 37235, USA}
\newcommand{\waseda}{Waseda University, Advanced Research Institute for Science and Engineering, 17 Kikui-cho, Shinjuku-ku, Tokyo 162-0044, Japan}
\newcommand{\weizmann}{Weizmann Institute, Rehovot 76100, Israel}
\newcommand{\wigner}{Institute for Particle and Nuclear Physics, Wigner Research Centre for Physics, Hungarian Academy of Sciences (Wigner RCP, RMKI) H-1525 Budapest 114, POBox 49, Budapest, Hungary}
\newcommand{\yonsei}{Yonsei University, IPAP, Seoul 120-749, Korea}
\newcommand{\zagreb}{University of Zagreb, Faculty of Science, Department of Physics, Bijeni\v{c}ka 32, HR-10002 Zagreb, Croatia}
\affiliation{\abilene}
\affiliation{\augie}
\affiliation{\banaras}
\affiliation{\barc}
\affiliation{\baruch}
\affiliation{\bnlcoll}
\affiliation{\bnlphys}
\affiliation{\caucr}
\affiliation{\charlesczech}
\affiliation{\chonbuk}
\affiliation{\ciae}
\affiliation{\cns}
\affiliation{\colorado}
\affiliation{\columbia}
\affiliation{\czechtech}
\affiliation{\dapnia}
\affiliation{\debrecen}
\affiliation{\elte}
\affiliation{\ewha}
\affiliation{\fit}
\affiliation{\fsu}
\affiliation{\gsu}
\affiliation{\hanyang}
\affiliation{\hiroshima}
\affiliation{\howard}
\affiliation{\ihepprot}
\affiliation{\illuiuc}
\affiliation{\inrras}
\affiliation{\instpasczech}
\affiliation{\isu}
\affiliation{\jaea}
\affiliation{\jinrdubna}
\affiliation{\jyvaskyla}
\affiliation{\kek}
\affiliation{\korea}
\affiliation{\kurchatov}
\affiliation{\kyoto}
\affiliation{\labllr}
\affiliation{\lahorelums}
\affiliation{\lawllnl}
\affiliation{\losalamos}
\affiliation{\lpc}
\affiliation{\lund}
\affiliation{\maryland}
\affiliation{\mass}
\affiliation{\michigan}
\affiliation{\muenster}
\affiliation{\muhlenberg}
\affiliation{\myongji}
\affiliation{\nagasaki}
\affiliation{\newmex}
\affiliation{\nmsu}
\affiliation{\ohio}
\affiliation{\ornl}
\affiliation{\orsay}
\affiliation{\peking}
\affiliation{\pnpi}
\affiliation{\riken}
\affiliation{\rikjrbrc}
\affiliation{\rikkyo}
\affiliation{\saispbstu}
\affiliation{\saopaulo}
\affiliation{\seoulnat}
\affiliation{\stonybrkc}
\affiliation{\stonycrkp}
\affiliation{\tenn}
\affiliation{\titech}
\affiliation{\tsukuba}
\affiliation{\vandy}
\affiliation{\waseda}
\affiliation{\weizmann}
\affiliation{\wigner}
\affiliation{\yonsei}
\affiliation{\zagreb}
\author{A.~Adare} \affiliation{\colorado}
\author{S.~Afanasiev} \affiliation{\jinrdubna}
\author{C.~Aidala} \affiliation{\mass} \affiliation{\michigan}
\author{N.N.~Ajitanand} \affiliation{\stonybrkc}
\author{Y.~Akiba} \affiliation{\riken} \affiliation{\rikjrbrc}
\author{R.~Akimoto} \affiliation{\cns}
\author{H.~Al-Bataineh} \affiliation{\nmsu}
\author{J.~Alexander} \affiliation{\stonybrkc}
\author{M.~Alfred} \affiliation{\howard}
\author{A.~Angerami} \affiliation{\columbia}
% ...
\collaboration{PHENIX Collaboration} \noaffiliation

\date{\today}

\title{Comparison of the space-time extent of the pion emission source in 
$d$$+$Au and Au$+$Au collisions at $\sqrt{s_{{NN}}}=200$ GeV}

%------------------------------------------------------------------------------|

\begin{abstract}

%\linenumbers

Two-pion interferometry measurements in $d$$+$Au and Au$+$Au collisions at 
$\sqrt{s_{{NN}}}=200$ GeV are used to extract and compare the Gaussian 
source radii R$_{{\rm out}}$, R$_{{\rm side}}$, and R$_{{\rm long}}$, 
which characterize the space-time extent of the emission sources. The 
comparisons, which are performed as a function of collision centrality and 
the mean transverse momentum for pion pairs, indicate strikingly similar 
patterns for the $d$$+$Au and Au$+$Au systems. They also indicate a linear 
dependence of R$_{{\rm side}}$ on the initial transverse geometric size 
$\bar{R}$, as well as a smaller freeze-out size for the $d$$+$Au system. 
These patterns point to the important role of final-state rescattering 
effects in the reaction dynamics of $d$$+$Au collisions.

\end{abstract}

\pacs{25.75.Dw} 
	
\maketitle

%\textbf{*** page break for PRL word count ***}  
%\clearpage

%----------------------------------
% Introduction
%----------------------------------
%

Recent measurements for hadrons emitted in $d$$+$Au collisions 
($\sqrt{s_{NN}}=200$~GeV) at the Relativistic Heavy Ion Collider  
\cite{phenix:ppg161,Adare:2013piz}, and in $p$$+$Pb collisions 
($\sqrt{s_{NN}}=5.02$~TeV) at the Large Hadron Collider  
\cite{CMS:2012qk,Abelev:2012ola,Aad:2012gla,Aad:2013fja,Chatrchyan:2013nka,ABELEV:2013wsa}, 
have indicated a surprising ridge structure in two-dimensional 
correlations in relative pseudorapidity ($\Delta\eta$) and azimuthal angle 
($\Delta\phi$). Elucidation of the origin of these long-range correlations 
should advance the current understanding of the very early-time dynamics 
of the matter produced in hadron nucleus ($p$$+A$ and $d$$+$$A$) and nucleus 
nucleus ($A$$+$$A$) collisions 
\cite{Bozek:2011if,Dusling:2012iga,Shuryak:2013ke,Lacey:2013eia}.

Two successful approaches are currently being employed to study long-range 
correlations. The Color Glass Condensate (CGC) approach accounts for these 
correlations via an enhancement of interference diagrams in the saturation 
regime \cite{Dusling:2013oia,Dusling:2012iga}. The viscous hydrodynamical 
approach 
\cite{Bozek:2011if,Bozek:2013uha,Bzdak:2013zma,Qin:2013bha,Werner:2013ipa,Bozek:2013ska} 
accounts for the same correlations via collective harmonic flow. Thus, it 
is presently not clear whether the long-range ridge, observed in $d$$+$Au 
and $p$$+$Pb collisions, stems from (i) the final-state effects inherent in 
a hydrodynamical description, (ii) the initial-state effects driven by the 
correlations of gluons already present in the nucleon and nuclear 
wave functions or (iii) an interplay between these two mechanisms.

Interferometry measurements of the space-time extent of the emitting 
sources produced in $A$$+$$A$ collisions indicate characteristic patterns (as 
a function of collision centrality and the mean transverse momentum $k_T$, 
of particle pairs) which serve as a ``fingerprint'' for collective 
expansion 
\cite{Lisa:2005dd,Adler:2004rq,Afanasiev:2007kk,Adams:2004yc,Aamodt:2011mr}. 
Thus, it might be expected that similar measurements for $d$$+$$A$ and $p$$+A$ 
collisions could provide an important avenue to independently constrain 
the role of final-state interactions in the reaction dynamics for these 
systems \cite{Bozek:2013df,Bzdak:2013zma}. An observed similarity between 
the characteristic patterns for the space-time extent of $A$$+$$A$ and $d$$+$$A$ 
(or $p$$+A$) collisions would give a strong indication for the importance 
of final-state rescattering effects in $d$$+$$A$ and $p$$+A$ collisions.

In this Letter, we use the interferometry technique of Hanbury Brown and 
Twiss (HBT) \cite{hbt_org} to perform detailed differential measurements 
of two-pion correlation functions 
\cite{Zajc:1984vb,Pratt:1984su,Ganz:1998zj,Adler:2004rq,Adams:2004yc,Lisa:2005dd, 
Afanasiev:2007kk,Aamodt:2011mr,Adare:2014vax} in $d$$+$Au and Au$+$Au 
collisions at $\sqrt{ s_{{NN}}} = 200$~GeV. In turn, these correlation 
functions are used to extract and study the HBT radii which characterize 
the space-time extent of the emission sources for the two systems. We find 
striking similarities in the detailed dependence of the HBT radii for both 
systems on collision centrality, transverse system-size, and $k_T$, which 
point to the importance of final state rescattering effects in the 
reaction dynamics of $d$$+$Au collisions.

%----------------------------------
% Detectors and Basic information
%----------------------------------

The present analysis uses the data recorded by the PHENIX experiment 
during 2007 and 2008. The collision vertex $z$ 
(along the beam axis) was constrained to $|z|< 30$~cm of the nominal 
crossing point. Collision centrality was determined from the charge 
distribution measured in the beam-beam counters, which span the 
pseudorapidity range $3.0<|\eta|<3.9$~\cite{Adare:2013nff}. Track and 
momentum reconstruction for charged particles were performed by combining 
hits from the drift chambers (DC) and pad chambers in the PHENIX 
central spectrometers ($|\eta|<0.35$). Charged pions were identified by 
combining time-of-flight from the time-of-flight detector and the 
electromagnetic calorimeters (EMCal) \cite{EMC} covering azimuthal angle 
$\Delta\phi < \pi/2$, with momentum reconstructed from the DC 
and pad-chamber hits 
in the magnetic field. Particles within 2 standard deviations of the peak 
for charged pions in the squared mass distribution were identified as 
pions for momenta up to $\sim$ 1 GeV/$c$ as detailed in 
Ref.~\cite{Adare:2013esx}.

%----------------------------------
% Analysis method for HBT
%----------------------------------

The two-pion correlation function is defined as the ratio $C_2\left({\bf 
q}\right)=A\left({\bf q}\right)/B\left({\bf q}\right)$, where $A\left({\bf 
q}\right)$ is the measured distribution of the relative momentum 
difference ${\bf q}={\bf p}_2-{\bf p}_1$ between particle pairs with 
momenta ${\bf p}_1$ and ${\bf p}_2$; $B\left({\bf q}\right)$ is the 
so-called background distribution, obtained from particle pairs in which 
each particle is selected from a different event but with similar event 
centralities, vertex positions, and charge sign. The relative momentum 
${\bf q}$ is calculated in the longitudinally co-moving system, where the 
longitudinal pair momentum is zero. It is also decomposed into its three 
components, ${q}_{{\rm out}}$, ${q}_{{\rm side}}$, and 
${q}_{{\rm long}}$, following the Bertsch--Pratt 
convention~\cite{Bertsch:1989vn,Pratt:1986cc}. That is, the ``out'' axis 
points along the pair transverse momentum, the ``side'' axis is 
perpendicular to the out axis in the transverse plane, and the ``long'' 
axis points along the beam.

Track merging and track splitting \cite{Adler:2004rq,Afanasiev:2007kk} 
were suppressed via pair selection cuts in the DC and the EMCal.  
%
% In the DC, pion pairs with $\Delta z \lt 5$~cm and $\Delta \phi \lt 
% 0.07$~rad, as well as pairs with $\Delta z \lt 70$~cm and 
% $\Delta \phi \lt  0.02$~rad$\!$\, were removed from the analysis.  
% The EMCal cuts were used to remove pairs with separation distance 
% $\Delta r \lt 17$~cm.
%
Correlation functions were studied as a function of collision centrality, 
as well as for different pion-pair transverse momenta 
$k_T = |{\bf p}_{T,1}+{\bf p}_{T,2}|/2$ or transverse mass 
$m_{T}=\sqrt{(k_{T}^2+m_{\pi}^2)}$, where $m_{\pi}$ is the pion mass.

%----------------------------------
% Fig.1
%----------------------------------
\begin{figure}[htb]
\includegraphics[width=1.0\linewidth]{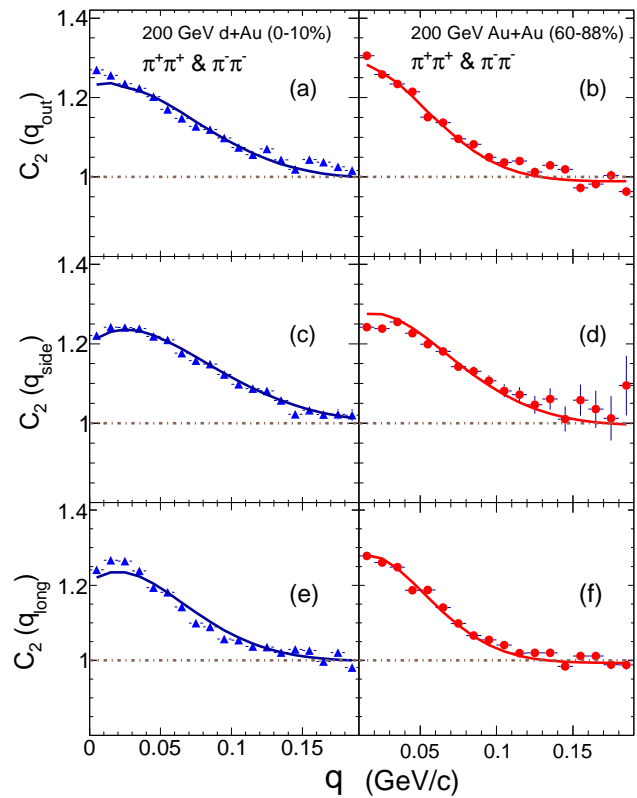}
\caption{(Color online) Slices of the three-dimensional two-pion 
($\pi^{+}\pi^{+}$ and $\pi^{-}\pi^{-}$) correlation functions for central 
$d$$+$Au collisions (left panels) and peripheral Au$+$Au collisions for 
$0.2 \lt k_T \lt 0.7$~GeV/$c$ ($\left< k_T \right> = 0.39$~GeV/$c$). These 
centrality selections give similar $N_{\rm part}$ values for the two 
systems. The curves represent fits to the correlation function (see text).
}
\label{fig1}
\end{figure}

Figure~\ref{fig1} shows a representative set of slices from the 
three-dimensional two-pion correlation functions for central (0\%--10\%) 
$d$$+$Au and peripheral (60\%--88\%) Au$+$Au collisions for $0.2 \lt k_T 
\lt 0.7$~GeV/$c$ ($\left< k_T \right> = 0.39$~GeV/$c$).  They all show the 
familiar Bose--Einstein enhancement peak at low $q$, as well as the 
expected difference in the peak widths for $d$$+$Au and Au$+$Au. The 
latter reflects the difference in the emission source sizes for the 
$d$$+$Au and Au$+$Au systems. Note that these centrality selections give 
similar values for the number of participants ($N_{\rm part} = 16.7 \pm 
1.1$ and $15.7 \pm 1.6$), but different values for the transverse 
geometric size ($\bar{R} =0.44 \pm 0.02$ fm and $0.71 \pm 0.06$ fm) 
for $d$$+$Au and Au$+$Au, respectively.

A similar set of correlation functions was extracted for several 
centralities to facilitate detailed comparisons of the $d$$+$Au and Au$+$Au 
emission sources as a function of $N_{\rm part}$, $\bar{R}$ and $k_T$. 
Monte Carlo Glauber (MC-Glauber) calculations 
\cite{glauber,Lacey:2010hw,Adare:2013nff} were used to compute 
$N_{\rm part}$ and $\bar{R}$ as a function of collision centrality, 
from the two-dimensional profile of the density of point-like sources in 
the transverse plane $\rho_s(\mathbf{r_{\perp}})$, where 
${1}/{\bar{R}}~=~\sqrt{\left({1}/{\sigma_x^2}+{1}/{\sigma_y^2}\right)}$, 
with $\sigma_x$ and $\sigma_y$ the respective root-mean-square widths of 
the density distributions \cite{Bhalerao:2005mm}. The systematic 
uncertainties for these geometric quantities, obtained via variation of 
the MC-Glauber model parameters, are less than 10\% \cite{Adare:2013nff}.

To aid the comparisons, the measured correlation functions were fitted 
with the following expression (in which cross-terms are assumed to be 
negligible) which accounts for the Bose--Einstein enhancement and the 
Coulomb interaction between pion pairs \cite{Bowler,Sinyukov:1998fc}:
\begin{eqnarray}
C_{2}({\bf q}) = N [ ( \lambda (1+G({\bf q})) ) F_{c} + (1-\lambda)], \nonumber \\ 
G({\bf q}) \cong \exp( -R_{{\rm side}}^{2}q_{{\rm side}}^{2} -R_{{\rm out}}^{2}q_{{\rm out}}^{2} -R_{{\rm long}}^{2}q_{{\rm long}}^{2}),
%-2R_{{\rm out-side}}^{2}q_{{\rm side}}q_{{\rm out}} )
%
\label{eq:sinyukov}
\end{eqnarray}
where $N$ is a normalization factor, $\lambda$ is the correlation 
strength, $F_{c}$ is the Coulomb correction factor \cite{Sinyukov:1998fc} 
evaluated with the Coulomb wave function, and ${R}_{{\rm out}}$, 
${R}_{{\rm side}}$, and ${R}_{{\rm long}}$ are the Gaussian HBT radii 
which characterize the emission source. ${R}_{{\rm side}}$ and 
${R}_{{\rm long}}$ are related to the transverse and longitudinal size of 
the source; ${R}_{{\rm out}}$ includes additional effects from the 
emission duration.

Excellent fits to the correlation functions for the $d$$+$Au and Au$+$Au 
systems were obtained and cross-checked to confirm agreement with our 
earlier measurements for Au$+$Au and $d$$+$Au collisions 
\cite{Adler:2004rq,Afanasiev:2007kk,Chung:2005ra}. The fit parameters for 
$\pi^+\pi^+$ and $\pi^-\pi^-$ pairs were also found to agree within 
statistical errors; the data for $\pi^+\pi^+$ and $\pi^-\pi^-$ were 
therefore combined. The systematic uncertainties for the fits were 
estimated via variations of the cuts used to generate the correlation 
functions (single track cuts, pair selection cuts and particle 
identification cuts). Typical values of the systematic uncertainties are 
5.0\%(7.5\%) for the extracted values of $R_{{\rm out}}$, 
$R_{{\rm side}}$, and $R_{{\rm long}}$ for Au$+$Au($d$$+$Au) and do not 
exceed 7.5\%(10.0\%).

%----------------------------------ral 
% Fig.2
%----------------------------------
%
\begin{figure}[htb]
\includegraphics[width=1.0\linewidth]{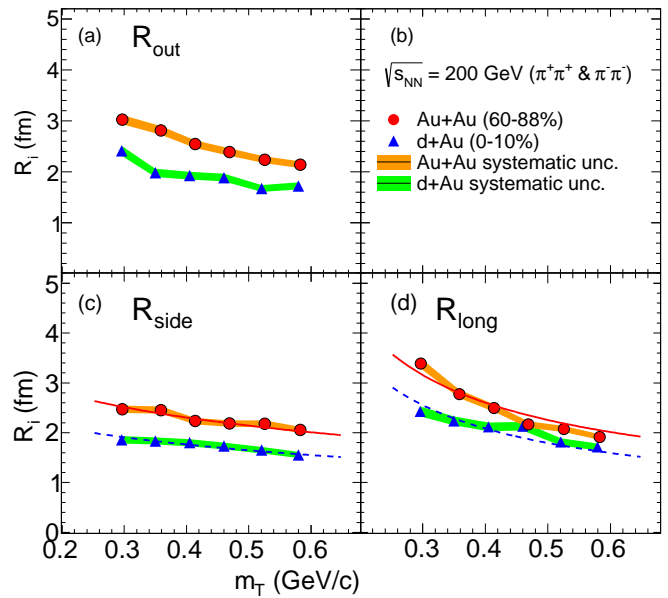}
\caption{(Color online) Comparison of the $m_{T}$ dependence of 
$R_{{\rm out}}$, $R_{{\rm side}}$, and $R_{{\rm long}}$ for 0\%--10\% 
central $d$$+$Au and 60\%--88\% central Au$+$Au collisions. The solid and dashed 
curves in panels (c) and (d) indicate fits to the data (see text). The 
color bands indicate the systematic uncertainties. }
\label{fig2}
\end{figure}

%----------------------------------
% Fig.3
%----------------------------------
%
\begin{figure}[htb]
\includegraphics[width=1.0\linewidth]{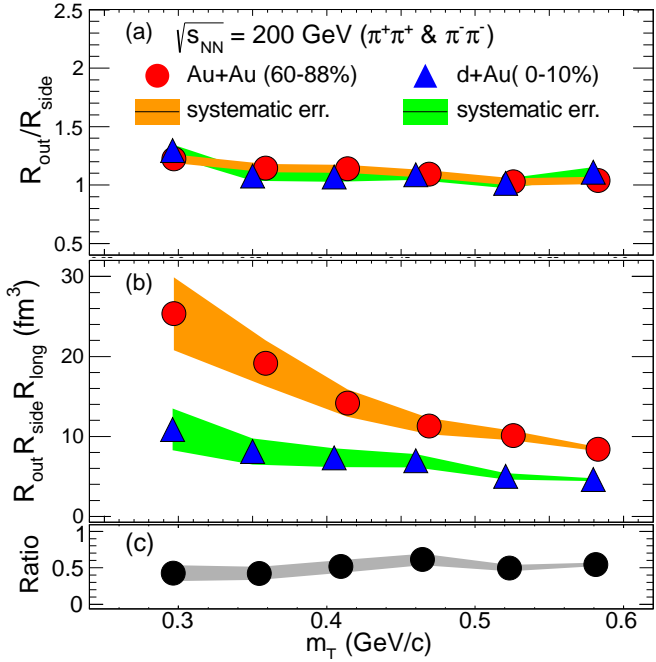}
\caption{(Color online) 
Comparison of the $m_T$ dependence of; (a) the ratio 
$R_{{\rm out}}/R_{{\rm side}}$; (b) the freeze-out volume, and (c) the 
ratio of the freeze-out volumes, for 0\%--10\% central $d$$+$Au and 60\%--88\% 
central Au$+$Au collisions.
}
\label{fig3}
\end{figure}

Figure~\ref{fig2} shows a comparison of the $m_{T}$ dependence of 
$R_{{\rm out}}$, $R_{{\rm side}}$, and $R_{{\rm long}}$ for 0\%--10\% 
central $d$$+$Au and 60\%--88\% central Au$+$Au collisions, i.e. similar 
values of $N_{\rm part}$.  The radii for $d$$+$Au and Au$+$Au show a 
decreasing trend with increasing values of $m_T$. The $R_{{\rm out}}$ 
radius is also comparable to $R_{{\rm side}}$ (for both systems) and the 
$m_T$ dependence of the ratio $R_{{\rm out}}/R_{{\rm side}}$ is flat or 
gently decreasing, as shown in Fig. \ref{fig3}(a). The same trends have 
been observed in central Au$+$Au and Pb$+$Pb collisions 
\cite{Ganz:1998zj,Adler:2004rq,Afanasiev:2007kk,Adams:2004yc,Aamodt:2011mr} 
and are commonly identified as a characteristic signature for the 
expansion of an emitting source of short emission duration, driven by 
final-state rescattering effects \cite{Herrmann:1994rr}. Therefore, we 
interpret the similarity between the observed patterns for Au$+$Au and 
$d$$+$Au in Figs.~\ref{fig2} and \ref{fig3}, as an indication for 
final-state rescattering effects in the reaction dynamics for $d$$+$Au.

The curves in Fig.~\ref{fig2} show blast wave expansion model inspired 
fits to $R_{{\rm side}}$ and $R_{{\rm long}}$ with fit functions 
\cite{Chapman:1995nz,Tomasik:2003gt}:
\begin{eqnarray}
R_{{\rm side}} = R_{{\rm geom}}/ \sqrt{(1+\beta^{2}(m_{T}/T))}, \\
R_{{\rm long}} = \tau_{0}\sqrt{(T/m_{T})[(K_{2}(m_{T}/T))/(K_{1}(m_{T}/T))]},
\end{eqnarray}
\begin{table}[htb]
\caption{Fit parameters} 
\begin{ruledtabular} \begin{tabular}{ccc}
  & $d$$+$Au  &   Au$+$Au   \\	\hline 
  $\tau_0$ (fm/$c$)	
& $3.2 \pm 0.04 \pm 0.4\:{\rm (syst)}$ 
& $3.8 \pm  0.04 \pm 0.3\:{\rm (syst)}$ \\
  $\chi^2/ndf$               
& 26/5  
& 24/5  \\  \\
  $R_{{\rm geom}}$ (fm)   
& $2.2 \pm 0.03 \pm 0.2\:{\rm (syst)}$ 
& $2.8 \pm 0.03 \pm 0.2\:{\rm (syst)}$  \\ 
  $\chi^2/ndf$              
& 6/5  
& 4/5  \\ 
\end{tabular} \end{ruledtabular}
\label{tab:fitParameters}
\end{table}
where $R_{{\rm geom}}$ is the geometrical radius at freeze-out and 
$\tau_{0}$ is the expansion time. The requisite freeze-out temperatures 
($T = 0.118 \pm 0.02$ and $0.123 \pm 0.02$~GeV) and expansion velocities 
($\left<\beta\right> = 0.42 \pm 0.03$ and $0.38 \pm 0.08$~$c$) for $d$$+$Au 
and Au$+$Au (respectively), are interpolated values obtained from a blast 
wave fit to the $p_{T}$ spectra for identified charged hadrons 
\cite{Abelev:2008ab}; $K_1$ and $K_2$ are modified Bessel functions. The 
fit results are summarized in Table \ref{tab:fitParameters};  they suggest 
a smaller transverse freeze-out size for the $d$$+$Au emitting source.

Figure~\ref{fig3}(b) further illustrates the difference via the 
$m_{T}$ dependence of the freeze-out volume, evaluated as the product 
$(R_{{\rm out}}\times R_{{\rm side}}\times R_{{\rm long}})$ for the 
same $\left< N_{\rm part}\right>$ values employed in Fig.~\ref{fig2}. 
The magnitudes of the freeze-out volumes for Au$+$Au are larger. However, 
within uncertainties, the fall-off with increasing $m_T$ is comparable for 
$d$$+$Au and Au$+$Au as shown by the ratio in Fig.~\ref{fig3}(c).

%----------------------------------
% Fig.4
%----------------------------------

\begin{figure}[t]
\includegraphics[width=1.0\linewidth]{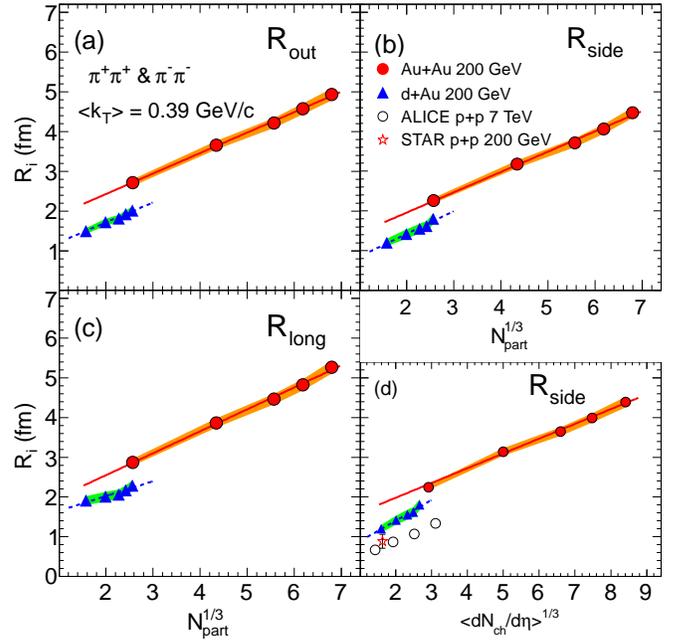}
\caption{(Color online) (a,b,c) HBT radii ($R_{{\rm out}}$, 
$R_{{\rm side}}$ and $R_{{\rm long}}$) vs. $N_{\rm part}^{1/3}$ for 
Au$+$Au and $d$$+$Au collisions.  (d) $R_{{\rm side}}$ vs 
${\langle dN_{\rm ch}/d\eta \rangle}^{1/3}$ 
for Au$+$Au, $d$$+$Au and $p$$+$$p$ \cite{Aggarwal:2010aa,Aamodt:2011kd} 
collisions.  Results are shown for $\left< k_{T} \right>$ = 0.39 GeV/$c$. 
The solid and dashed curves represent linear fits to the Au$+$Au and 
$d$$+$Au data, respectively.  The color bands indicate the systematic 
uncertainties.
}
\label{fig4}
\end{figure}

Detailed comparisons were also made as a function of collision centrality. 
Figs.~\ref{fig4}(a-c) show one such comparison of $R_{{\rm out}}$, 
$R_{{\rm side}}$, and $R_{{\rm long}}$ for $d$$+$Au and Au$+$Au, as a 
function of $N_{\rm part}^{1/3}$ for $\left< k_{T} \right>$ = 0.39 
GeV/$c$. The solid and dashed curves represent linear fits to the Au$+$Au 
and $d$$+$Au data, respectively.  The data for $R_{{\rm out}}$ and 
$R_{{\rm side}}$ indicate a similar linear increase with 
$N_{\rm part}^{1/3}$, albeit with larger magnitudes for Au$+$Au. An 
apparent slope difference between $d$$+$Au and Au$+$Au for $R_{{\rm long}}$ 
(Fig.~\ref{fig4}c), could be the result of a difference in the 
longitudinal dynamics for the two systems.  The representative plot of 
$R_{{\rm side}}$ vs. $(dN/d\eta)^{1/3}$ shown in Fig.~\ref{fig4}(d), 
indicates that the HBT radii for $d$$+$Au do follow the linear dependence 
previously observed for $A$+$A$ and $p$+$p$ collisions 
\cite{Aamodt:2011kd}, but with separate magnitudes for each system.

The dependencies shown in Figs.~\ref{fig4}(a-c) suggest that the pattern 
of a strong correlation between the transverse freeze-out size and the 
initial geometric size, is similar for both $d$$+$Au and Au$+$Au. They also 
suggest that at $\sqrt{s_{{NN}}} = 200$~GeV, the change in the transverse 
expansion rates with centrality (defined by $N_{\rm part}$) is similar 
for central $d$$+$Au and peripheral Au$+$Au collisions.

%----------------------------------
% Fig.5
%----------------------------------

\begin{figure*}[t]
\includegraphics[width=0.7\linewidth]{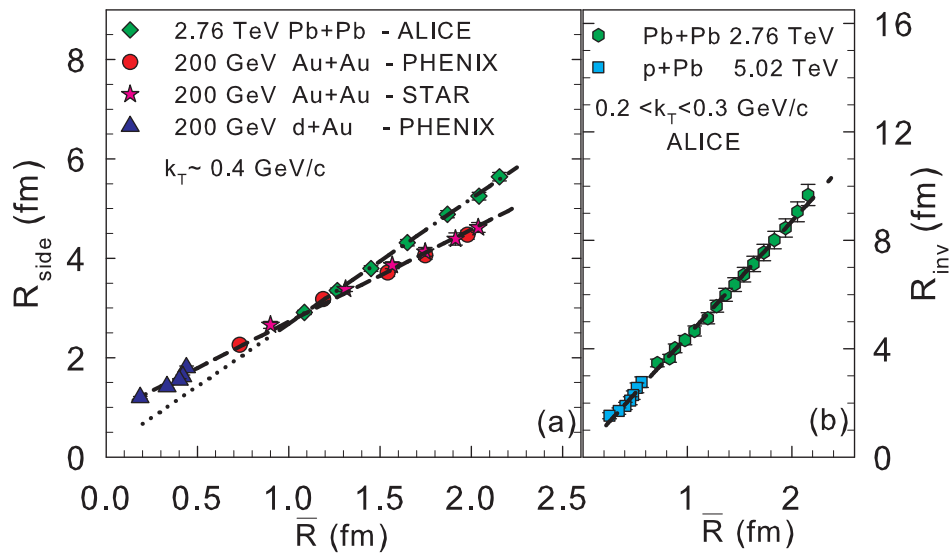}
\caption{(Color online) 
(a) $R_{{\rm side}}$ vs. $\bar{R}$, for $\left< k_{T} \right> \approx 
0.4$~GeV/$c$ for $d$$+$Au, Au$+$Au and Pb$+$Pb collisions as indicated. 
(b) $R_{{\rm inv}}$ vs. $\bar{R}$ for $p$$+$Pb and Pb$+$Pb collisions.
The ALICE and STAR data are taken from Refs. \cite{Aamodt:2011mr,Abelev:2014pja} and 
\cite{Adams:2004yc} respectively. Systematic uncertainties are 
5.0\%(7.5\%) for Au$+$Au($d$$+$Au). The dashed curves in $a$ ($b$) 
represents a linear fit to the Au$+$Au and $d$$+$Au ($p$$+$Pb and Pb$+$Pb) data sets. 
The dotted curve is an extrapolation of the dashed-dot curve.
} 
\label{fig5}
\end{figure*}

%----------------------------------

%----------------------------------

In some models~\cite{Lacey:2013is,Lacey:2013qua,Shuryak:2013ke}, the 
expansion time is proportional to the initial geometric size 
$\tau \propto \bar{R}$.  Therefore, $\bar{R}$ might be expected to be a 
more natural scaling variable for the HBT radii of expanding systems. The 
detailed dependencies of $R_{{\rm side}}$ on $\bar{R}$ are compared in 
Fig.~\ref{fig5}(a) for $d$$+$Au and Au$+$Au collisions at $\sqrt{s_{{NN}}} 
= 200$~GeV, and Pb$+$Pb collisions at $\sqrt{s_{{NN}}} = 2.76$~TeV for 
$\left< k_{T} \right> \sim 0.4$~GeV/$c$. Fig.~\ref{fig5}(b) shows a 
similar dependence for recent $R_{{\rm inv}}$ measurements for $p$$+$Pb 
and Pb$+$Pb collisions \cite{Abelev:2014pja}. The comparisons indicate 
that $R_{{\rm side}}$ and $R_{{\rm inv}}$ scale linearly with $\bar{R}$ 
for all of these systems. This pattern is consistent with the observed 
$1/\bar{R}$ scaling of collective anisotropic flow 
\cite{Lacey:2013eia,Lacey:2013is}. The dashed curves in 
Figs.~\ref{fig5}(a) and (b) are linear fits to the $d$$+$Au and Au$+$Au 
($p$$+$Pb and Pb$+$Pb) data sets; they suggest similar slopes for $d$$+$Au 
and Au$+$Au ($p$$+$Pb and Pb$+$Pb). The fit to the Pb$+$Pb data in 
Figs.~\ref{fig5}(a) (dotted curve) indicates a larger slope for Pb$+$Pb 
collisions at the much higher energy of $\sqrt{s_{{NN}}} = 2.76$~TeV, 
where the expansion rate is expected to be larger. The observed 
dependencies of $R_{{\rm side}}$ and $R_{{\rm inv}}$ on $\bar{R}$ 
reinforce our earlier inferences that the final-state rescattering 
effects, which are known to play a dominant role in Au$+$Au and Pb$+$Pb 
collisions, also play an important role in $d$$+$Au and $p$$+$Pb 
collisions.

%----------------------------------
% Summary
%----------------------------------

% In summary, we have presented detailed comparisons of the HBT 
% radii $R_{{\rm out}}$, $R_{{\rm side}}$, and $R_{{\rm long}}$, 
% extracted from two-pion interferometry measurements for 
% $d$$+$Au and Au$+$Au collisions at $\sqrt{s_{{NN}}}= 200$~GeV. 
% The comparisons, which emphasize trends (as a function of $k_T$ 
% and centrality) commonly associated with hydrodynamic-like 
% collective expansion, indicate excellent agreement between the 
% respective patterns for the $d$$+$Au and Au$+$Au systems.
%
In summary, we have presented detailed comparisons of HBT radii, which 
emphasize trends commonly associated with hydrodynamic-like expansion. 
Excellent agreement is found between the patterns for the $d$$+$Au and 
Au$+$Au systems. The radii extracted for the two systems at similar 
$\left< N_{\rm part} \right>$ show similar dependencies on $m_T$, which 
indicate a smaller geometric size (at freeze-out) for the emitting source 
in $d$$+$Au collisions. The $R_{{\rm side}}$ and $R_{{\rm inv}}$ radii for 
different systems show scaling with the initial transverse size $\bar{R}$, 
across several collision energies, which is indicative of 
hydrodynamic-like collective expansion driven by final-state rescattering 
effects. An investigation of the interesting possibility of a similar 
pattern in high multiplicity $p$+$p$ events is deferred to a future study. 
Our present findings, which support the view that the expansion dynamics 
for $d$$+$Au and Au$+$Au ($p$$+$Pb and Pb$+$Pb) are similar, constitute a 
significant contribution towards a more comprehensive understanding of the 
very early-time dynamics of the matter produced in hadronic collisions.

%\textbf{*** page break for PRL word count $<$3.5 pages $<$7 columns} 
%\clearpage

%%%%%%%%%%%%%%%%%%%%%%%%%  Acknowledgments 

\section*{ACKNOWLEDGMENTS}  % Run-7, 8, 13 long form for all journals

We thank the staff of the Collider-Accelerator and Physics
Departments at Brookhaven National Laboratory and the staff of
the other PHENIX participating institutions for their vital
contributions.  We acknowledge support from the 
Office of Nuclear Physics in the
Office of Science of the Department of Energy, the
National Science Foundation, 
Abilene Christian University Research Council, 
Research Foundation of SUNY, and 
Dean of the College of Arts and Sciences, Vanderbilt University (U.S.A),
Ministry of Education, Culture, Sports, Science, and Technology
and the Japan Society for the Promotion of Science (Japan),
Conselho Nacional de Desenvolvimento Cient\'{\i}fico e
Tecnol{\'o}gico and Funda\c c{\~a}o de Amparo {\`a} Pesquisa do
Estado de S{\~a}o Paulo (Brazil),
Natural Science Foundation of China (P.~R.~China),
Ministry of Education, Youth and Sports (Czech Republic),
Centre National de la Recherche Scientifique, Commissariat
{\`a} l'{\'E}nergie Atomique, and Institut National de Physique
Nucl{\'e}aire et de Physique des Particules (France),
Bundesministerium f\"ur Bildung und Forschung, Deutscher
Akademischer Austausch Dienst, and Alexander von Humboldt Stiftung (Germany),
Hungarian National Science Fund, OTKA (Hungary), 
Department of Atomic Energy and Department of Science and Technology (India), 
Israel Science Foundation (Israel), 
National Research Foundation of Korea of the Ministry of Science,
ICT, and Future Planning (Korea),
Physics Department, Lahore University of Management Sciences (Pakistan),
Ministry of Education and Science, Russian Academy of Sciences,
Federal Agency of Atomic Energy (Russia),
VR and Wallenberg Foundation (Sweden), 
the U.S. Civilian Research and Development Foundation for the
Independent States of the Former Soviet Union, 
the US-Hungarian Fulbright Foundation for Educational Exchange,
and the US-Israel Binational Science Foundation.

%\clearpage
%%%%%%%%%%%%%%%%%%%%%%%%%%%  References 

%\bibliography{ppg164x0}   

\end{document}